\documentclass[12pt]{iopart}
\usepackage{graphicx}
\usepackage{epstopdf}
\usepackage{hyperref}
\usepackage{cite}
\hypersetup{
	colorlinks=true,
	linkcolor=red,
	filecolor=red,
	citecolor = red,      
	urlcolor=red,
}

\usepackage{iopams}  
\begin{document}

\title[]{Coriolis contribution to excited states of odd-mass nuclei with different deformation-dependent mass coefficients}

\author{A. Ait Ben Hammou, M. Oulne}

\address{High Energy Physics and Astrophysics Laboratory, FSSM, \\
	Cadi Ayyad University, Marrakesh, 40000, Morocco}
\ead{abderrahim.aitbenhammou@edu.uca.ac.ma and oulne@uca.ma}
\vspace{10pt}
\begin{indented}
\item[]
\end{indented}

\begin{abstract}
Within the collective Bohr Hamiltonian, the adoption of a mass tensor as a function of  collective coordinates has demonstrated its importance for describing the structure of nuclei. On the other hand, for odd-mass  nuclei, the Coriolis interaction between the rotational and single-particle motions affects significantly the structure of nuclear excited states. In the framework of a recently developed extended Bohr Hamiltonian, by considering  the Deformation-Dependent Mass Formalism whith different mass parameters for the rotation and the two $ \beta $ and $ \gamma $ vibrations  and taking into account the Coriolis contribution, we investigate the bands structure of the $^{173}$Yb, $^{163}$Dy, $^{155}$Eu and $^{153}$Eu nuclei. Excited-state energies and B(E2) transition probabilities are calculated and compared with the available experimental data. Besides, we investigate the effect of DDMF and the Coriolis force on nuclear observables.\\\\

\end{abstract}

%

\section{Introduction}

The Bohr Hamiltonian \cite{1} has been proved to be among the most suitable ways for studying the collective structure in heavy nuclei \cite{2,3,4,5,6,7}. A number of analytical solutions with different current model  potentials have been proposed in this framework using a constant mass parameter for all the vibration and rotation modes\cite{8,9,10,11,12,13,14}. However,  further developments, by several authors, have been elaborated within new approaches like for example: the Deformation Dependent Mass  Formalism (DDMF) \cite{15,16,17,18,19}, where the tensor mass is not considered as a constant but depends on collective coordinate, and the procedure of different mass coefficients, where to each collective motion mode corresponds a mass coefficient \cite{20,21,22,23}. This latter has been  proved to be important for describing properties of nuclei, especially interband E2 transition probabilities. As to the DDMF, in addition to allowing to improve numerical results particularly for energy ratios, it  reduces the increase rate of the moment of inertia, removing a main drawback of the adopted model  and leading to an improved agreement with the corresponding experimental data \cite{24}. As an additional new approach, we can find, the introduction of the minimal length formalism \cite{25,26}, inspired by Heisenberg algebra, which allows to improve the precision of  numerical calculations of physical observables, in particular the energy spectra.

The theory of collective quadrupole-type excitations, which takes into account the coupling of single-particle and collective motions, enables us to explain some regularities observed in the excitation spectra of deformed odd-mass nuclei. Recently, some studies are realized within the DDMF for some of such nuclei \cite{27} but without taking into account the Coriolis interaction. They have been realized in a simple way where quantum numbers of the projections of angular momentum of a nucleus to the third principal axis $ K $ and that of the external nucleon $ \Omega $ are good quantum numbers.  In  \cite{28}, using different mass parameters, the authors have proved that the Coriolis interaction pushes, for both neutron-odd nuclei $^{163}$Dy and $^{173}$Yb, the energies upward in all bands. Such an effect was observed to be stronger in the $ \beta $ and $ \gamma $ bands than in the ground-state ($ i\equiv g.s.$) one. As a consequence, the positions of the $ \beta $ and $ \gamma $ bands change relatively in respect to that of the $g.s.$. On the other side, such an effect has only a small impact on the E2 transition probabilities by decreasing them. 

A similar study to that of \cite{28}  has been carried out in \cite{29} with the aim of investigating the band structure of proton-odd nuclei $^{155}$Eu and $^{153}$Eu. However, the used energy formula in both \cite{28} and \cite{29} was inaccurate as proved in \cite{17} for even-even nuclei and in \cite{27} for the case of odd-A nuclei.  

In this present work we extend the previous one of \cite{27} by incuding Coriolis interaction of the angular momentum of the external nucleon with that of the core of the odd-mass nucleus. In this case, $ K $ and $ \Omega $ are nonconserved quantities. Then, the value of angular momentum $ j $ of the external nucleon directly contributes to excited-state energies and wave functions. Then, we study the excited states of the odd nuclei $^{173}$Yb, $^{163}$Dy, $^{155}$Eu and $^{153}$Eu. Moreover, we will focus on the effect of the introduction of DDMF and Coriolis interaction in the calculations.

This paper is organized as follows. In \sref{disc2} we propose a Bohr Hamiltonian including the Coriolis contribution with three different mass coefficients. In \sref{disc3} we give the Bohr Hamiltonian built with a mass depending on the deformation coordinate $ \beta $. In \sref{disc4} we present the expressions we have obtained for  energy levels, wave functions, and reduced E2 transition probabilities, considering $ K $ and $ \Omega $ as nonconserved quantities and using three different mass parameters for rotation, $ \beta $ and $ \gamma $ vibrations in the framework of DDMF. In \sref{disc5} are made analysis and comparison of obtained theoretical results with experimental data for spectra as well as electromagnetic transition rates. Finally, \sref{disc6} is devoted to the conclusion.

\section{Bohr Hamiltonian with different mass coefficients}\label{disc2}

For small amplitudes of $ \gamma $ vibration around $ \gamma=0 $ and $ \beta $ vibration around $ \beta = \beta_{0}\neq 0 $, the Schr\"{o}dinger equation, in the case of an odd-mass nucleus, can be represented as \cite{6} 
\begin{eqnarray}
(H_{vib}+H_{rot}+H_{p}(x)+H_{int})\Psi=E\Psi,\label{eq1}
\end{eqnarray}
where, by including different mass parameters \cite{22}, 
\begin{eqnarray}
 H_{vib}=-\frac{\hbar^2}{2}\Big(\frac{1}{B_{ \beta}}\frac{\partial^2}{\partial\beta^2}+\frac{2}{B_{ \gamma}}\frac{1}{ \beta}\frac{\partial}{\partial\beta}+\frac{2}{B_{ \beta}}\frac{1}{ \beta}\frac{\partial}{\partial\beta}+\frac{1}{B_{ \gamma}\beta^2}\frac{1}{ \gamma}\frac{\partial}{\partial\gamma}(\gamma\frac{\partial}{\partial\gamma})
 \nonumber\\
  \qquad \quad \, -\frac{1}{B_{\gamma}}\frac{1}{4\beta^2}(\frac{1}{\gamma^2}+\frac{1}{3})(L_{3}-j_{3})^2\Big)
 +V(\beta,\gamma)\label{eq2}
\end{eqnarray}
is the operator describing the $ \beta $ and $ \gamma $ vibrations of the nuclear core surface,
\begin{eqnarray}
H_{rot}=\frac{\hbar ^2}{6B_{rot}\beta^2}\Big[(L^2+j^2-L_{3}^2-j_{3}^2-2(L_{1}j_{1}+L_{2}j_{2})\Big]
\label{eq3}
\end{eqnarray}
is the nuclear rotational energy operator that includes the Coriolis interaction represented by the component $ 2(L_{1}j_{1}+L_{2}j_{2}) $ and $ H_{p}(x)+H_{int} $ is the operator of the Hamiltonian of the external nucleon in the nucleus core field. $ H_{p}(x)$ takes into account the central-symmetrical part of the field.
\begin{eqnarray}
H_{int}=-\beta<T>\Big(3j_{3}^2-j^2\Big)\label{eq4} 
 \end{eqnarray}
is the operator taking into account the non-spherical part of the nuclear core field. $ B_{\beta}$, $B_{\gamma}$, and $B_{rot}$ are the mass parameters. $ L $ is the total angular momentum of the nucleus, $L_{1}$,$L_{2}$, and $L_{3}$ are its projections on the principal axes of the nucleus. $ j $, $j_{1}$, $j_{2}$, and $j_{3}$ are the operator of a single nucleon external to the core, and its projections, and $ x $ are the space and spin nucleon coordinates. $ T (r) $ is a function of the distance between
the single nucleon and the center of the nuclear core \cite{6,7}. $ \langle T \rangle $ is its average value over internal states of the external nucleon, assuming zero nuclear surface oscillations.

\section{Bohr Hamiltonian with different deformation-dependent mass parameters}\label{disc3}

The Schr\"{o}dinger equation corresponding to the Bohr Hamiltonian with  three
deformation-dependent mass coefficients is given by \cite{17}
\begin{eqnarray}
\fl \frac{\hbar^2}{2\langle i|B_{0}|i\rangle}( \frac{-\sqrt{f}}{\beta^4}\frac{\partial}{\partial\beta}\beta^4 f \frac{\partial}{\partial\beta}\sqrt{f}-\frac{f^2}{\beta^2 sin3\gamma} \frac{\partial}{\partial\gamma}sin3\gamma \frac{\partial}{\partial\gamma} + 
\frac{f^2}{4\beta^2}\sum _{k=1,2,3}\frac{(L_{k}-j_{k})^2}{sin^2(\gamma-\frac{2}{3}\pi k)})\Psi\nonumber\\-f^2\beta\langle T\rangle(3j_{3}^2-j^2)\Psi+V_{eff}\Psi=E\Psi , 
\label{eq5}
\end{eqnarray} 
with 
\begin{eqnarray}
\fl V_{eff}=V(\beta,\gamma) + \frac{\hbar^2}{2\langle i|B_{0}|i\rangle}(\frac{1}{2}(1-\delta-\lambda)f\nabla^2f +(\frac{1}{2}-\delta)(\frac{1}{2}-\lambda)(\nabla f)^2), \label{eq6}
\end{eqnarray}
where $ f $ is the deformation function. It depends only on the radial coordinate $ \beta $. Then, only the $ \beta $ part of the resulting equation will be affected. $ \langle i|B_{0}|i\rangle $ defines the mass parameters $B_{rot}$, $ B_{\beta}$ and $B_{\gamma}$ for $g.s.$ ($ i\equiv g.s.$), $\beta$-vibrational state ($ i\equiv \beta$) and $\gamma$-vibrational state ($ i\equiv \gamma$),respectively \cite{20}. Originated from the construction procedure of the kinetic
energy term within DDMF, $ \delta $ and $ \lambda $ are free parameters \cite{15}. 

\section{Energy  levels, wave functions and E2 transition probabilities}\label{disc4}

In order to solve equation \eref{eq5}, the separation of the variables $\beta$ and $\gamma$ is obtained with the potential chosen as in \cite{8,10} and \cite{17,18,19}:
\begin{eqnarray}
V(\beta,\gamma)=U(\beta)+\frac{f^2}{\beta^2}W(\gamma),
\label{eq7}
\end{eqnarray}
where 
\begin{eqnarray}
U(\beta)=V_{0}\Big(\frac{\beta}{\beta_{0}}-\frac{\beta_{0}}{\beta}\Big)^2
\label{eq8}
\end{eqnarray}
is the Davidson potential for $ \beta $-part, and
\begin{eqnarray}
W(\gamma)=\frac{1}{2}(\beta_{0}^4C_{\gamma})\gamma^2
\label{eq9}
\end{eqnarray}
is the harmonic oscillator potential for $ \gamma $-part.
$\beta_{0}$ is the position of the minimum of the potential in $\beta$, $ C_{\gamma} $ is a free parameter, and $ V_{0} $ represents the depth of the minimum, located at $\beta_{0}$.

For the deformation function, we use  the following special form:
\begin{eqnarray}
f=1+a\beta^2,\qquad a<<1
\label{eq10}
\end{eqnarray}
where $ a $ is the deformation parameter.

The resolution of equation \eref{eq5}, using the Asymptotic Iteration Method (AIM) \cite{30}, gives  
\begin{eqnarray}
E_{n_{\beta}n_{\gamma}L|m|\tau}=\frac{\hbar^2}{2B_{\beta}}\Big(K_{0}+\frac{a}{2}(2+\frac{B_{\beta}}{B_{\gamma}}+2p+2q+pq)\nonumber\\ \qquad \qquad \qquad +2a(2+p+q)n_{\beta}+4an^2_{\beta}\Big)+\epsilon_{p},
\label{eq11}
\end{eqnarray}
where
\begin{eqnarray}
\eqalign{
q\equiv q^\tau_{n_{\gamma}}(L,|m|)=\sqrt{1+4K_{-2}} \cr p\equiv p^\tau_{n_{\gamma}}(L,|m|)=\sqrt{4\frac{B_{\beta}}{B_{\gamma}}-3+4\frac{K_{2}}{a^2}}},
\label{eq12}
\end{eqnarray}
and
\begin{eqnarray}
\eqalign{K_2=\frac{a^2}{2}\Big[\Big(1+\frac{B_{\beta}}{B_{\gamma}}\Big)\Big(6\frac{B_{\beta}}{B_{\gamma}}+(1-2\delta)(1-2\lambda)\cr\qquad \, +5(1-\delta-\lambda)\Big)+\frac{2B_{\beta}}{\hbar^2}\Lambda\Big]+\frac{2g_{\beta}}{\beta^4_{0}},
\cr
K_0=\frac{a}{2}\Big[\Big(1+\frac{B_{\beta}}{B_{\gamma}}\Big)\Big(8\frac{B_{\beta}}{B_{\gamma}}+5(1-\delta-\lambda)\Big)+\frac{4B_{\beta}}{\hbar^2}\Lambda\Big]-\frac{4g_{\beta}}{\beta^2_{0}},\cr
K_{-2}=\frac{B_{\beta}}{B_{\gamma}}\Big(1+\frac{B_{\beta}}{B_{\gamma}}\Big)+\frac{B_{\beta}}{\hbar^2}\Lambda+2g_{\beta}}
\label{eq13}
\end{eqnarray}
where $g_{\beta}=\frac{B_{\beta}V_{0}\beta_{0}^2}{\hbar^2}$, $ \tau $ distinguishes different
states of the same $ L $, $ n_{\beta} $ is the quantum number of $\beta$-vibrations, and $ \epsilon_{p} $ is the corresponding energy of $ H_{p}$ which determines the distance between the single-particle spherical orbits.

The eigenvalues of the $\gamma$-vibrational part of the Hamiltonian plus the term of the rotational energy are determined by the following expression:
\begin{eqnarray}
\frac{B_{\beta}}{\hbar^2}\Lambda=\frac{2}{g}\frac{B_{\beta}}{B_{\gamma}}(1+2n_{\gamma}+|m|)
+\frac{m^2}{3}\frac{B_{\beta}}{B_{\gamma}}+\varepsilon_{|m|L\tau},
\label{eq14}
\end{eqnarray}
where  $g=\frac{1}{\beta_{0}^2}\frac{\hbar}{\sqrt{B_{\gamma}C_{\gamma}}}$ and $n_{\gamma}$ is the quantum number of $ \gamma $-vibrations. The values of $ m $ are connected with $ K $ and $ \Omega $ through the condition $ K - \Omega = 2m $, where $ m $ should be an integer. 

The following determinant is calculated in order to determine eigenvalues
and eigenfunction of the rotational part of the Hamiltonian:
\begin{eqnarray}
||\langle LjKm|X| LjK'm'\rangle-\varepsilon_{|m|L\tau}\delta_{KK'}\delta_{mm'}||=0, \label{eq15}
\end{eqnarray}
where
\begin{eqnarray}
X=\frac{1}{3}\frac{B_{\beta}}{B_{rot}}\Big[L(L+1)+j(j+1)-L_{3}^2-j_{3}^2\nonumber\\ \qquad 
-2(L_{1}j_{1}+L_{2}j_{2})\Big]-\frac{1}{3\xi}\Big[3j_{3}^2-j(j+1)\Big], \label{eq16}
\end{eqnarray}
and $\xi=\frac{\hbar^2}{6B_{\beta}\beta_{0}^3\langle T\rangle}$. Because of the quantum numbers
$ K $ and $ \Omega $ which are not conserved in this present considerations, not only  the diagonal elements of the Hamiltonian but also nondiagonal ones do contribute to the energies and
E2 transition probabilities. The diagonal elements are given as follows:
\begin{eqnarray}
\fl \langle LjKm|X| LjKm\rangle=\frac{1}{3}\frac{B_{\beta}}{B_{rot}}\Big[L(L+1)+j(j+1)-K^2-(K-2m)^2-(-1)^{L-j}\nonumber\\ \qquad \quad
\times (L+1/2)(j+1/2)\delta_{K1/2}\delta_{m0}\Big]-\frac{1}{3\xi}\Big[3(K-2m)^2-j(j+1)\Big].
\label{eq17}
\end{eqnarray}
The nondiagonal elements are
\begin{eqnarray}
\langle LjKm|X|LjK\pm m\rangle=\frac{1}{3}\frac{B_{\beta}}{B_{rot}}\Big[(L\mp K)(L\pm K+1)\Big]^{1/2} \nonumber\\ \qquad \qquad \qquad \qquad \qquad \times \Big[(j\mp K\pm 2m)(j\pm K\mp 2m+1)\Big]^{1/2} .
\label{eq18}
\end{eqnarray}

The bands are classified by the quantum numbers $n_{\beta} $, $ n_{\gamma} $ and $ m $, such as the $g.s.$ band with $ n_{\beta}=0 $, $ n_{\gamma}=0 $, $ m=0 $; the $\beta $-band with $ n_{\beta}=1$ ,$ n_{\gamma}=0 $, $ m=0 $; and the $ \gamma $-band with $ n_{\beta}=0$, $ n_{\gamma}=0 $ ,$ m=1 $.

The corresponding wave function is
\begin{eqnarray}
\Psi=\beta^{-(1+B_{\beta}/B_{\gamma})}R_{n_{\beta},L}(\beta)\sum _{{m}{K}}A_{LK}^{m\tau}\chi_{n_{\gamma}|m|}(\gamma)|LMjKm\rangle,
\label{eq19}
\end{eqnarray}
where
\begin{eqnarray}
R_{n_{\beta},L}(\beta)=\beta^{\frac{1}{2}(1+q)}(1+a\beta^2)^{-n_{\beta}-\frac{1}{2}(1+\frac{B_{\beta}}{B_{\gamma}})-\frac{1}{4}(p+q)}\phi(\beta),
\label{eq20}
\end{eqnarray}
with
\begin{eqnarray}
\phi(\beta)=N_{n_{\beta}}~_{~ 2}F_1(-n_{\beta},-n_{\beta}-\frac{p}{2};-2n_{\beta}-\frac{(q+p)}{2};1+a\beta^2),
\label{eq21}
\end{eqnarray}
\begin{eqnarray}
\chi_{n_{\gamma},|m|}(\gamma)=N_{n_{\gamma},|m|}\gamma^{|m|}e^{-\frac{\gamma^2}{2g}}~_{~ 1}F_1(-n_{\gamma},1+|m|,\frac{\gamma^2}{g}),
\label{eq22}
\end{eqnarray}
and
\begin{eqnarray}
\fl |LMjKm\rangle=\sqrt{\frac{2L+1}{16\pi^2}}\Big[D^L_{MK}(\theta_{i})\varphi^j_{K-2m}(x)
+(-1)^{L-j}D^L_{M-K}(\theta_{i})\varphi^j_{-K+2m}(x)\Big],\nonumber\\ \qquad \quad i=1,2 \: or \: 3.
\label{eq23}
\end{eqnarray}

Here, $ _{~ 2}F_1(-n_{\beta},-n_{\beta}-\frac{p}{2};-2n_{\beta}-\frac{(q+p)}{2};1+a\beta^2) $ and $ _{~ 1}F_1(-n_{\gamma},1+|m|,\frac{\gamma^2}{g}) $ are hypergeometrical functions, $ D(\theta_{i}) $ is the Wigner function, $ \varphi(x) $ is the wave function of  single-particle state, $ A_{LK}^{m\tau} $ are the coefficients of the expansion of the wave function \cite{31}, and $ N_{n_{\beta}} $ with $ N_{n_{\gamma},|m|} $ are normalization constants of radial and angular wave functions, respectively.

According to the relationship between hypergeometric functions and generalized Jacobi polynomials on one side (equation (4.22.1) of \cite{32}), and between hypergeometric functions and the Laguerre polynomials on the other side, we have
\begin{eqnarray}
\fl R(t)=N_{n_{\beta}}2^{-(1+\frac{B_{\beta}}{B_{\gamma}})/2-(q+p)/4}a^{-(1+q)/4}(1-t)^{(1+2\frac{B_{\beta}}{B_{\gamma}}+p)/4}(1+t)^{(q+1)/4}P^{(q/2,p/2)}_{n_{\beta}}(t),\nonumber\\ \qquad \qquad
t=\frac{a\beta^2-1}{a\beta^2+1}
\label{diseqn}
\end{eqnarray}
and
\begin{eqnarray}
\chi_{n_{\gamma},|m|}(\gamma)=N_{n_{\gamma},|m|}\gamma^{|m|}e^{-\frac{\gamma^2}{2g}}L_{n_{\gamma}}^{|m|}(\frac{\gamma^2}{g}),
\label{diseqn}
\end{eqnarray}
with 
\begin{eqnarray}
\fl N_{n_{\beta}}=(2a^{q/2+1}n_{\beta}!)^{1/2}\Bigg[\frac{\Gamma(n_{\beta}+\frac{q+p}{2}+1)\Gamma(2n_{\beta}+\frac{q+p}{2}+1+\frac{B_{\beta}}{B_{\gamma}})}{\Gamma(n_{\beta}+\frac{q}{2}+1)\Gamma(n_{\beta}+\frac{B_{\beta}}{B_{\gamma}}+\frac{p}{2})\Gamma(2n_{\beta}+\frac{q+p}{2}+1)}\Bigg]^{1/2},
\label{diseqn}
\end{eqnarray}
and
\begin{eqnarray}
N_{n_{\gamma},|m|}=\Big[\frac{2}{3}\frac{n_{\gamma}!}{g^{1+|m|}\Gamma(|m|+n_{\gamma})}\Big]^{1/2},
\label{diseqn}
\end{eqnarray}

B(E2) transition rate, taking into account the nonconservation of $ K $ \cite{28}, is given by
\begin{eqnarray}
B(E2;n_{\beta}Ln_{\gamma}K|m|\longrightarrow n'_{\beta}L'n'_{\gamma}K'|m'|)=
\frac{5t^2}{16\pi}\Big|\sum _{{K}{K'}}A_{LK}^{m\tau}A_{L'K'}^{m'\tau'}G\Big|^2 \nonumber\\ \qquad \qquad
\qquad \qquad \qquad \qquad \qquad \qquad \quad
\times I^2_{n_{\beta}L,n'_{\beta}L'}C^2_{n_{\gamma},|m|,n'_{\gamma},|m'|},
\label{diseqn}
\end{eqnarray}
where $ t $ is a scaling factor, $ G=<L,K,2,K'-K|L',K'>$ is the Clebsch-Gordan coefficient dictating the angular momentum selection rules, while $ I_{n_{\beta}L,n'_{\beta}L'} $ and $ C_{n_{\gamma},|m|,n'_{\gamma},|m'|} $ are integrals over the shape variables $ \beta $ and $ \gamma $ with integration measures \cite{17}, 
\begin{eqnarray}
I_{n_{\beta}L,n'_{\beta}L'}=\int \beta R_{n_{\beta},L}(\beta)R_{n'_{\beta},L'}(\beta) \,\mathrm{d\beta},
\label{diseqn}
\end{eqnarray}

\begin{eqnarray}
C_{n_{\gamma},|m|,n'_{\gamma},|m'|}=\int sin \gamma\chi_{n_{\gamma}|m|}\chi_{n'_{\gamma}|m'|}|sin 3\gamma|\mathrm{d\gamma}.
\label{eq30}
\end{eqnarray}

\section{Numerical results and discussions }\label{disc5}

The $ 5/2^-[512] $, $ 5/2^-[523] $ and $ 5/2^+[413] $ ground states of $^{173}$Yb, $^{163}$Dy and the Eu isotopes, respectively, are determined from the Nilsson model\cite{33}. For both $^{173}$Yb and $^{163}$Dy, the Shell-model calculations predict that the orbital of greatest influence in the $g.s.$ structure is the neutron $ f_{7/2} $ orbital \cite{28}. And for both $^{155}$Eu and $^{153}$Eu, they predict that the contribution of the proton $ g_{7/2} $ orbital is dominant in comparison with other orbitals \cite{29}. $ \epsilon_{p} $ has the same value for each $ K $ band which  originates from spherical orbital. 

When $ m = 0 $, the angular momentum vector of the prolate core is perpendicular to the axis of symmetry, and thus it cannot contribute to the value of $ K $. Then, the value of $ K $ is determined by the projection of the angular momentum of the last nucleon \cite{6,7}. Its possible values should be up to $ j = 7/2 $ which is, here, the appropriate value for the angular momentum of the external nucleon. It can be noted here, as we can see in equation \eref{eq17}, that only when $ K = 1/2 $ and $ m = 0 $ does the Coriolis interaction affect the diagonal elements of the rotational part of the Hamilonian.

The calculation is performed for two cases: one out of the DDMF and the other within it. The first case corresponds to the value $ a=0 $ and it gives a correction to the previous energy formula used in  \cite{28,29}. In this case, we determine the values of the free parameters $g$, $g_{\beta}$, $B_{\beta}/B_{rot}$ and $B_{\beta}/B_{\gamma}$ using the correct formula:
\begin{eqnarray}
E_{n_{\beta}n_{\gamma}L|m|\tau}=\sqrt{2\frac{V_{0}^2}{g_{\beta}}}\Big(1+2n_{\beta}+\frac{1}{2}q^\tau_{n_{\gamma}}(L,|m|)-\sqrt{2g_{\beta}}\Big)+\epsilon_{p},
\label{diseqn}
\end{eqnarray}
with
\begin{eqnarray}
\frac{1}{2}q^\tau_{n_{\gamma}}(L,|m|)=\sqrt{\frac{1}{4}+\frac{B_{\beta}}{B_{\gamma}}\Big(1+\frac{B_{\beta}}{B_{\gamma}}\Big)+\frac{B_{\beta}}{\hbar^2}\Lambda+2g_{\beta}},
\label{diseqn}
\end{eqnarray}
by adjusting the parameters in order to reproduce the optimal experimental spectrum. For this aim, we have used the root-mean-square (r.m.s) coefficient which describes the average deviation between theoretical predictions and experimental data:
\begin{eqnarray}
\sigma=\sqrt{\frac{\sum_ {i=1}^{n}(E_i(exp)-E_i(th))^2}{(n-1)E(7/2_{g.s.})^2}},     
\label{eq33}
\end{eqnarray}
where $ E_i(exp) $ is the experimental energy of the $ i^{th} $ level, $ E_i(th) $ the corresponding theoretical value, $ n $ the maximum number of considered levels and $ E(7/2_{g.s.}) $ the band head energy  under consideration. This coefficient is also used as a qualitative test of the agreement between the theoretical results and the experimental data. The obtained values of the parameters used in the calculation are given in \tref{tab1}.

\begin{table}
	\caption{\label{tab1}The values of the parameters used in calculations.} 
	
	\begin{indented}
		\lineup
		\item[]\begin{tabular}{@{}*{6}{lccccc}}
			\br                              
			$ Nucleus $ & $ \xi $ & $ g $ & $ g_{\beta} $ & $ B_{\beta}/B_{rot} $ & $ B_{\beta}/B_{\gamma} $\cr 
			\mr
			$ $$ ^{173} $$ $Yb &  0.0378  &  0.0607  &  382  &  5.42  &  6.48  \cr
			$ $$ ^{163} $$ $Dy &  0.0284  &  0.3607  &  770  &  7.83  &  5.45  \cr 
			$ $$ ^{155} $$ $Eu &  0.0337  &  0.4001  &  769  &  6.30  &  5.36  \cr 
			$ $$ ^{153} $$ $Eu &  0.0498  &  0.3895  &  274  &  6.30  &  5.21  \cr 
			\br
		\end{tabular}
	\end{indented}
\end{table}

In order to improve the obtained numerical results,  in the case where the Coriolis effect is taken into account, we have recalculated the energy ratios in the context of the DDMF and  determined its action on excited-state energies and E2 transition probabilities. So, the energy formula contains two supplementary parameters, namely: $ a $ and $ \beta_{0} $. The optimal values of both parameters are evaluated through  r.m.s fits of energy levels by making use of equation \eref{eq33} for each band of each nucleus. Note that, in the calculations of the second case, the values of the free parameters
$ \delta $ and $ \lambda $ are null as in \cite{15,17,19}.
 
\begin{table}
	\caption{\label{tab2}The comparison of the theoretical predictions of $E(L_{g.s.})$ normalized to the energy of the first excited state $E(7/2_{g.s.}^-)$ in two cases: $ a=0 $ and $ a\neq0 $ for $^{173}$Yb and $^{163}$Dy with those from \cite{28} and experimental values of \cite{34}.} 
	
	\begin{indented}
		\lineup
		\item[]\begin{tabular}{@{}*{10}{lccccccccc}}
			\br                              
			& \multicolumn{4}{c}{$ $$ ^{173} $$ $Yb}  &  & \multicolumn{4}{c}{$ $$ ^{163} $$ $Dy}\cr     
			\cline{2-5} \cline{7-10}\cr
			$ $$ L_{g.s.} $$ $ &\cite{28}& $ a=0 $ &   $ a\neq0 $  & $ Exp. $&  &\cite{28}& $ a=0 $ &   $ a\neq0 $  & $ Exp. $\cr
			\mr 
			$ 9/2^{-} $  & 2.38  & 2.29  & 2.29  & 2.28  & & 2.27 &2.29  & 2.29  & 2.28 \cr 
			$ 11/2^{-} $ & 3.82  & 3.82  & 3.83  & 3.84  & & 3.77 &3.82  & 3.82  & 3.84 \cr 
			$ 13/2^{-} $ & 6.04  & 5.68  & 5.72  & 5.67  & & 5.55 &5.70  & 5.72  & 5.66 \cr 
			$ 15/2^{-} $ & 7.68  & 7.65  & 7.72  & 7.77  & & 7.43 &7.65  & 7.68  & 7.75 \cr 
			$ 17/2^{-} $ & 10.99 & 10.14 & 10.27 & 10.13 & & 9.71 &10.20 & 10.25 & 10.13 \cr 
			$ 19/2^{-} $ & 12.62 & 12.41 & 12.61 & 12.75 & &11.84 &12.40 & 12.49 & 12.68 \cr 
			$ 21/2^{-} $ & 17.18 & 15.62 & 15.94 & 15.61 & &14.63 &15.74 & 15.89 & 15.49 \cr 
			$ 23/2^{-} $ & 18.69 & 18.02 & 18.44 & 18.72 & &16.87 &18.01 & 18.19 & 18.56 \cr 
			$ 25/2^{-} $ & 24.49 & 22.07 & 22.71 & 22.09 & &20.18 &22.29 & 22.59 & 21.82 \cr 
			$ 27/2^{-} $ & 25.83 & 24.41 & 25.20 & 25.67 & &22.42 &24.44 & 24.79 & 25.36 \cr
			\mr
			$ \sigma $   & 1.009 & 0.497 & 0.306 &       & &1.330 & 0.413& 0.374 & \cr
			$ a $        &       &       & 0.0679&       & &      &      & 0.0179& \cr
			$ \beta_{0} $ &      &       & 1.95  &       & &      &      & 2.69  & \cr
			\br
		\end{tabular}
	\end{indented}
\end{table}

\begin{table}
	\caption{\label{tab3}The comparison of the theoretical predictions of $E(L_{g.s.})$ normalized to the energy of the first excited state $E(7/2_{g.s.}^+)$ in two cases: $ a=0 $ and $ a\neq0 $ for $^{155}$Eu and $^{153}$Eu with those from \cite{29} and experimental values of \cite{34}.} 
	
	\begin{indented}
		\lineup
		\item[]\begin{tabular}{@{}*{10}{lccccccccc}}
			\br                              
			& \multicolumn{4}{c}{$ $$ ^{155} $$ $Eu}  &  & \multicolumn{4}{c}{$ $$ ^{153} $$ $Eu}\cr     
			\cline{2-5} \cline{7-10}\cr
			$ $$ L_{g.s.} $$ $ &\cite{29}& $ a=0 $ &   $ a\neq0 $  & $ Exp. $&  &\cite{29}& $ a=0 $ &   $ a\neq0 $  & $ Exp. $\cr
			\mr 
			$ 9/2^{+} $  & 2.28  & 2.29  & 2.28  & 2.28  & & 2.32 &2.32  & 2.32  & 2.32 \cr 
			$ 11/2^{+} $ & 3.76  & 3.82  & 3.80  & 3.83  & & 3.77 &3.76  & 3.78  & 3.90 \cr
			$ 13/2^{+} $ & 5.57  & 5.70  & 5.65  & 5.64  & & 5.74 &5.74  & 5.79  & 5.77 \cr
			$ 15/2^{+} $ & 7.41  & 7.68  & 7.58  & 7.69  & & 7.43 &7.41  & 7.49  & 7.85 \cr 
			$ 17/2^{+} $ & 9.75  & 10.22 & 10.06 & 9.99  & & 9.71 &10.24 & 10.38 & 10.21 \cr 
			$ 19/2^{+} $ & 11.79 & 12.49 & 12.27 & 12.50 & &12.90 &11.86 & 12.06 & 12.73 \cr 
			$ 21/2^{+} $ & 15.72 & 15.80 & 15.48 & 15.24 & &16.97 &15.73 & 16.08 & 15.51 \cr
			$ 23/2^{+} $ & 17.77 & 18.18 & 17.80 & 18.16 & &18.34 &17.13 & 17.54 & 18.40 \cr 
			$ 25/2^{+} $ & 21.59 & 22.42 & 21.92 & 21.28 & &23.45 &22.12 & 22.83 & 21.56 \cr 
			$ 27/2^{+} $ & 23.43 & 24.74 & 24.17 & 24.54 & &24.48 &23.21 & 23.99 & 24.77 \cr
			\mr
			$ \sigma $   & 0.512 & 0.437 & 0.299 &       & &0.818 &0.774 & 0.659 & \cr
			$ a $        &       &       & 0.0653 &     &  &       &     & 0.1011& \cr
			$ \beta_{0} $ &      &       & 7.87    &   &   &       &     & 1.24  & \cr
			\br
		\end{tabular}
	\end{indented}
\end{table}

\begin{table}
	\caption{\label{tab4}The comparison of the theoretical predictions of $E(L_{\beta})$ normalized to the energy of the first excited state $E(7/2_{g.s.}^-)$ in two cases: $ a=0 $ and $ a\neq0 $ for $^{173}$Yb and $^{163}$Dy with those from \cite{28} and experimental values of \cite{34}.} 
	
	\begin{indented}
		\lineup
		\item[]\begin{tabular}{@{}*{10}{lccccccccc}}
			\br                              
			& \multicolumn{4}{c}{$ $$ ^{173} $$ $Yb}  &  & \multicolumn{4}{c}{$ $$ ^{163} $$ $Dy}\cr     
			\cline{2-5} \cline{7-10}\cr
			$ $$ L_{\beta} $$ $ &\cite{28}& $ a=0 $ &   $ a\neq0 $  & $ Exp. $&  &\cite{28}& $ a=0 $ &   $ a\neq0 $  & $ Exp. $\cr
			\mr 
			$ 5/2^{-} $  & 11.78 & 11.41 & 11.36 & 11.78 & & 9.70  & 10.00 & 9.93 &  9.70  \cr 
			$ 7/2^{-} $  & 12.78 & 12.41 & 12.36 &       & & 10.70 & 11.00 & 10.93 & 10.91 \cr
			$ 9/2^{-} $  & 14.16 & 13.70 & 13.65 & 13.46 & & 11.98 & 12.29 & 12.22 & 12.47 \cr 
			$ 11/2^{-} $ & 15.60 & 15.23 & 15.18 & 14.75 & & 13.47 & 13.81 & 13.75 &   \cr 
			$ 13/2^{-} $ & 17.83 & 17.09 & 17.04 & 16.38 & & 15.25 & 15.70 & 15.63 &   \cr 
			$ 15/2^{-} $ & 19.46 & 19.06 & 19.02 & 19.48 & & 17.13 & 17.64 & 17.58 &   \cr 
			$ 17/2^{-} $ & 22.78 & 21.55 & 21.51 & 20.73 & & 19.42 & 20.19 & 20.13 &   \cr 
			$ 19/2^{-} $ & 24.40 & 23.83 & 23.78 & 23.27 & & 21.55 & 22.40 & 22.34 &   \cr 
			$ 21/2^{-} $ & 28.96 & 27.04 & 27.00 & 27.99 & & 24.33 & 25.74 & 25.69 &   \cr 
			$ 23/2^{-} $ & 30.47 & 29.43 & 29.39 & 30.63 & & 26.58 & 28.00 & 27.95 &   \cr
			$ 25/2^{-} $ & 36.27 & 33.48 & 33.45 & 33.15 & & 29.89 & 32.29 & 32.25 &   \cr
			\mr
			$ \sigma $   & 1.274 & 0.712 & 0.710 &       & &0.377  & 0.253 & 0.239 & \cr
			$ a $        &       &       & 0.0010 &     & &       &        & 0.0018& \cr
			$ \beta_{0} $ &     &        & 2.00   &     & &       &        & 1.98  & \cr
			\br
		\end{tabular}
	\end{indented}
\end{table}
	
\begin{table}
	\caption{\label{tab5}The comparison of the theoretical predictions of $E(L_{\beta})$ normalized to the energy of the first excited state $E(7/2_{g.s.}^+)$ in two cases: $ a=0 $ and $ a\neq0 $ for $^{155}$Eu and $^{153}$Eu with those from \cite{29} and experimental values of  \cite{34}.} 
	
	\begin{indented}
		\lineup
		\item[]\begin{tabular}{@{}*{10}{lccccccccc}}
			\br                              
			& \multicolumn{4}{c}{$ $$ ^{155} $$ $Eu}  &  & \multicolumn{4}{c}{$ $$ ^{153} $$ $Eu}\cr     
			\cline{2-5} \cline{7-10}\cr
			$ $$ L_{\beta} $$ $ & \cite{29} & $ a=0 $ &   $ a\neq0 $  & $ Exp. $&  &\cite{29}& $ a=0 $ &   $ a\neq0 $  & $ Exp. $\cr
			\mr 
			$ 5/2^{+} $  & 14.98 & 12.33 & 12.33 & 12.46 &    & 12.93 & 8.13  & 7.79  & 7.41  \cr 
			$ 7/2^{+} $  & 15.58 & 13.33 & 13.33 & 13.41 &    & 13.54 & 9.13  & 8.79  & 8.78  \cr 
			$ 9/2^{+} $  & 16.34 & 14.62 & 14.62 & 14.65 &    & 14.34 & 10.44 & 10.11 & 10.51 \cr 
			$ 11/2^{+} $ & 17.23 & 16.15 & 16.15 &       &    & 15.22 & 11.89 & 11.57 &   \cr 
			$ 13/2^{+} $ & 18.31 & 18.03 & 18.03 &       &    & 16.41 & 13.87 & 13.56 &   \cr 
			$ 15/2^{+} $ & 19.42 & 20.01 & 20.01 &       &    & 17.40 & 15,54 & 15.25 &   \cr 
			$ 17/2^{+} $ & 20.85 & 22.54 & 22.54 &       &    & 19.10 & 18.37 & 18.11 &   \cr 
			$ 19/2^{+} $ & 22.06 & 24.82 & 24.82 &       &    & 20.04 & 19.99 & 19.75 &   \cr 
			$ 21/2^{+} $ & 23.89 & 28.13 & 28.13 &       &    & 22.34 & 23.86 & 23.68 &   \cr 
			$ 23/2^{+} $ & 25.09 & 30.51 & 30.51 &       &    & 23.12 & 25.25 & 25.09 &   \cr
			$ 25/2^{+} $ & 27.37 & 34.75 & 34.75 &       &    & 26.05 & 30.25 & 30.19 &   \cr
			\mr
			$ \sigma $   & 2.744  & 0.114 & 0.114 &       &    &5.82   & 0.566 & 0.388 & \cr
			$ a $        &        &  & $ 1.71\times10^{-11} $ & &  & &         & 0.1071& \cr
			$ \beta_{0} $ &     &         & 6.33 &        &     &    &         & 2.00  & \cr
			\br
		\end{tabular}
	\end{indented}
\end{table}
	
	The comparison of the calculated values, in both cases: without DDMF ($a=0$) and with DDMF ($a\neq0$), of  energy ratios $ E(L_{g.s})/E(7/2_{g.s.}) $ with experimental data \cite{34} and the results of Ermamatov et al. \cite{28,29} are given in \tref{tab2} for $^{173}$Yb and $^{163}$Dy and in \tref{tab3} for $^{155}$Eu and $^{153}$Eu. It shows that our results for $a=0$ agree globally with experimental data more than those of Ermamatov et al. \cite{28,29} ($ \sigma $ for our results is lower). On the other hand, it shows that there is a minor increase in energy when the DDMF is applied. Such an increase becomes larger as the level's angular momentum  increases too. This effect brings results slightly closer to the experimental values. Also note that these results (regarding those of \cite{27}), show that the DDMF affects more the energy levels of the $g.s.$ band when the Coriolis effect is taken into account with different mass parameters.
	
	The comparison of the calculated values of the energy ratios $ E(L_{\beta})/E(7/2_{g.s.}) $ in both cases: without DDMF ($a=0$) and with DDMF ($a\neq0$), with experimental data \cite{34} and the results of Ermamatov et al. \cite{28,29} are given in \tref{tab4} for $^{173}$Yb and $^{163}$Dy and in \tref{tab5} for $^{155}$Eu and $^{153}$Eu. We can also see that the obtained values for $a=0$ are generally more accurate than the results of Ermamatov et al. \cite{28,29}. Actually, these results would have been obtained by Ermamatov et al. if their formula in \cite{28,29} had been adequate. The introduction of DDMF slightly decreases the energies. It is the same effect when the Coriolis interaction is not taken into account \cite{27}. However, in the case of $^{155}$Eu, the $\beta$-band is not sensitive to the DDMF effect. 
	
	In \tref{tab6}, we compare our results for energy ratios $ E(L_{\gamma})/E(7/2_{g.s.}^-) $ for $^{173}$Yb, in both above-cited cases, with experimental data \cite{34} and the data from \cite{28}. It comes out that our results, for $a=0$, are in a good agreement with the experiment, but slightly lower than those of \cite{27} due to the Coriolis force. Moreover, no effect is observed when DDMF is applied as it has been also observed in \cite{27} when the mass coefficients are different. However, in the case where these mass coefficients are taken equal, the DDMF effect  reappears again in the $\gamma$-band in \cite{27}. So, within the mass coefficients procedure, only the $g.s.$ and $\beta$ bands are sensitive to the mass deformation.
	
\begin{figure}
	\centering
	\includegraphics[width=0.8\linewidth]{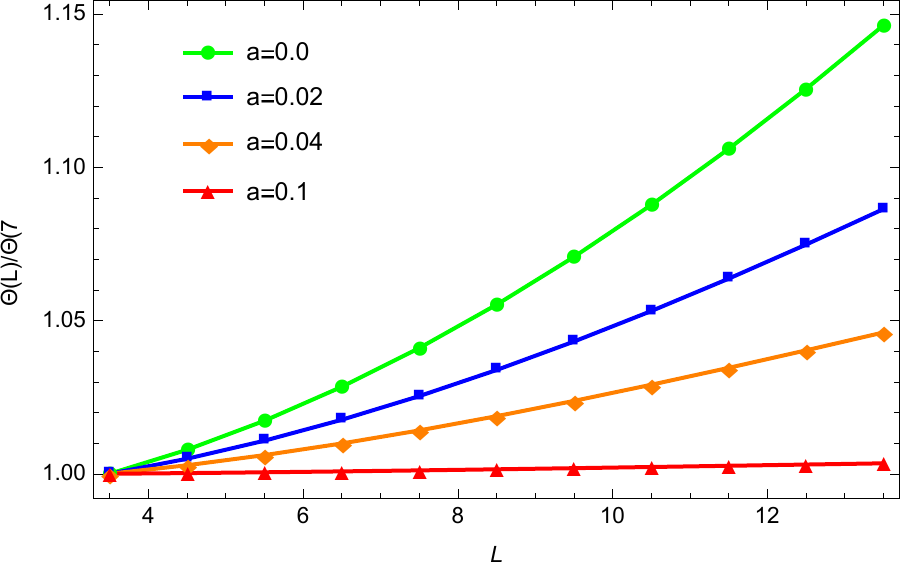}
	\caption{\label{fig1} The moments of inertia $\Theta(L)$ for the ground-state band, given by Equation \eref{eq34} and normalized to $\Theta(7/2)$, are shown for varying parameter $a$ without Coriolis interaction.}
\end{figure}

\begin{figure}
	\centering
	\includegraphics[width=0.8\linewidth]{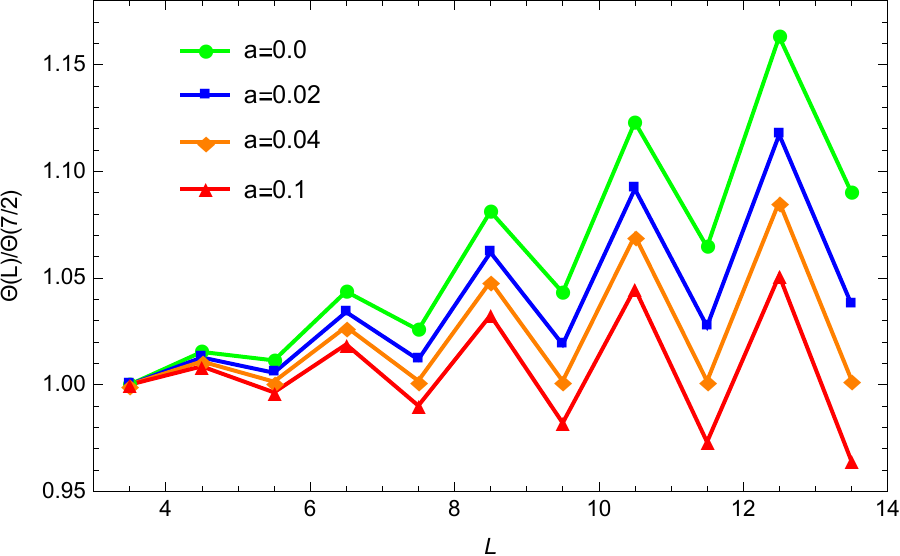}
	\caption{\label{fig2} The moments of inertia $\Theta(L)$ for the ground-state band, given by Equation \eref{eq34} and normalized to $\Theta(7/2)$, are shown for varying parameter $a$ with Coriolis interaction.}
\end{figure}

\begin{figure}
	\centering
	\includegraphics[width=0.8\linewidth]{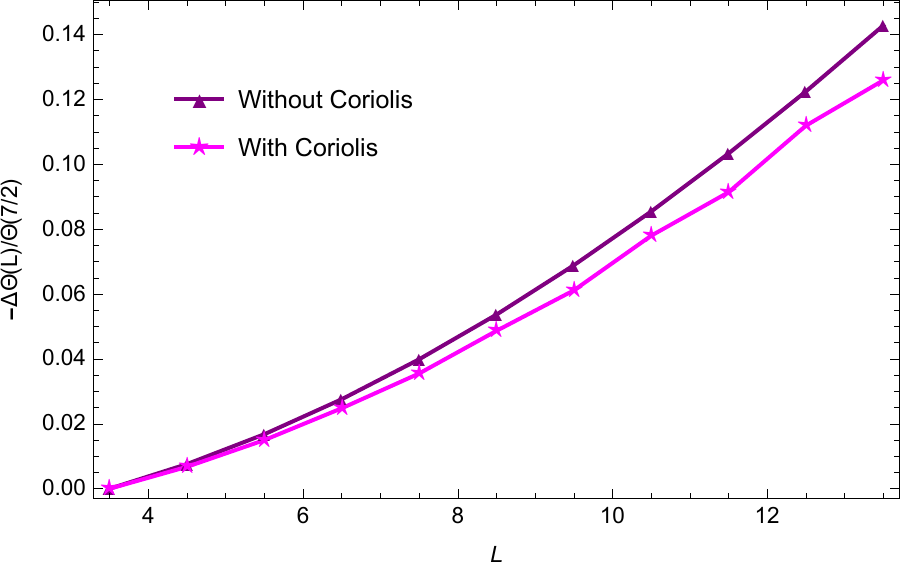}
	\caption{\label{fig3} Comparaison of the effect of DDMF to the Moments of inertia $\Theta(L)$ for the ground-state band normalized to $\Theta(7/2)$, with and without  Coriolis interaction for $a=0.1$}
\end{figure}

	In addition to the energy ratios, we have studied the effect of the DDMF on the moments of inertia of nuclei. Therefore, such an effect can be seen, in both cases: without and with Coriolis interaction, in \Fref{fig1} and \Fref{fig2}, respectively. The moment of inertia \cite{35} for the $g.s.$-band is given by	
	\begin{eqnarray}
	\Theta(L)=\frac{2L+1}{E(L+1)-E(L-1)},\label{eq34}
	\end{eqnarray}	
	normalized to $ E(7/2) $. In \Fref{fig1}, it is shown for different arbitrary values of the mass deformation parameter $ a $ for $^{173}$Yb nucleus as an example. In this case, it is apparent that the rate of increase of the moment of inertia with $ L $, observed for $ a = 0 $, is gradually moderated by increasing $ a $. In \Fref{fig2}, in addition to the effect of the DDMF parameter $ a $, we observe a staggering effect which is related to the Coriolis interaction. The amplitude of such an additional effect increases with the angular momentum $ L $, but the value of the inertial moment is kept slown down by the DDMF effect . In \Fref{fig3}, we compare the effect of the DDMF on the moments of inertia with and without  Coriolis interaction for $ a=0.1 $. The effect is reduced in the case where the Coriolis contribution is introduced. So, it appears that there is some opposite correlation between both effects: the DDMF effect decreases the moments of inertia, while the Coriolis interaction increases them. So, thanks to the DDMF, the effect of Coriolis interaction is dumped by DDMF effect as it is clearly shown in \Fref{fig2}.

\begin{table}
	\caption{\label{tab6} The comparison of the theoretical predictions of $E(L_{\gamma})$ normalized to the energy of the first excited state $E(7/2_{g.s.}^-)$ in two cases: $ a=0 $ and $ a\neq0 $ for $^{173}$Yb  with those from \cite{28} and experimental values of \cite{34}.} 
	
	\begin{indented}
		\lineup
		\item[]\begin{tabular}{@{}*{5}{lcccc}}
			\br                              
			$ $$ L_{\gamma} $$ $&\cite{28}&$ a=0 $&$ a\neq0 $&$ Exp. $\cr
			\mr 
			$ 9/2^{-} $  & 18.40 & 18.56 & 18.56 & 18.59 \cr 
			$ 11/2^{-} $ & 20.56 & 20.00 & 20.00 & 20.61 \cr 
			$ 13/2^{-} $ & 23.09 & 21.68 & 21.68 & 22.22 \cr 
			$ 15/2^{-} $ & 25.98 & 23.61 & 23.61 & 23.40 \cr 
			$ 17/2^{-} $ & 29.23 & 25.77 & 25.77 & 25.85 \cr 
			$ 19/2^{-} $ & 32.82 & 28.15 & 28.15 & 28.56 \cr 
			$ 21/2^{-} $ & 36.73 & 30.76 & 30.76 & 30.44 \cr 
			$ 23/2^{-} $ & 40.96 & 33.52 & 33.52 & 32.31 \cr
			\mr
			$ \sigma $   & 4.651 & 0.592 & 0.592 &   \cr
			$ a $        &       &       & 2.62$ \times10^{-6} $& \cr
			$ \beta _{0} $ &     &       & 0.16  &       \cr
			\br
		\end{tabular}
	\end{indented}
\end{table}
	
	The calculated values of reduced E2 intraband transition probabilities are given for the  $g.s.$-band  in units of  $B(E_{2};9/2_{g.s.} \rightarrow 5/2_{g.s.})$ in tables \ref{tab7}, \ref{tab8} and \ref{tab9} for $^{173}$Yb, $^{163}$Dy and $^{153}$Eu, respectively. They are compared with available experimental data of  \cite{34}, \cite{36}, \cite{37} and \cite{38}. These transitions correspond to the case where $ \Delta m=0 $. Then, the $\gamma $-integral part (equation \eref{eq30}) reduces to the orthonormality condition of the $\gamma $-wave functions: $ C_{n_{\gamma},|m|,n'_{\gamma},|m'|}=\delta_{n_{\gamma},n'_{\gamma}}\delta_{|m|,|m'|} $. Within  DDMF, we have used the same optimal values of the two parameters $a$ and $\beta _{0}$, previously obtained for the energy ratios in the $g.s.$-band. It is clearly shown that the inclusion of the DDMF, taking into account the Coriolis effect, similarly to the case of energy levels, decreases the E2 transition probabilities. However, this behavior becomes clearer as the angular momentum increases too. Note that this effect of DDMF on the E2 transition probabilities was not observed when the Coriolis effect is neglected \cite{27}.
	  
\begin{table}
	\caption{\label{tab7} The comparison of the theoretical predictions of $B(E_{2};L'_{g.s.} \rightarrow L_{g.s.})$ in units of $B(E_{2};9/2^-_{g.s.} \rightarrow 5/2^-_{g.s.})$ in two cases: $ a=0 $ and $ a\neq0 $ for $^{173}$Yb with those from \cite{28} and experimental values of \cite{34} and \cite{36}.} 
	
	\begin{indented}
		\lineup
		\item[]\begin{tabular}{@{}*{6}{cccccc}}
			\br                              
			$ $ $ L'_{g.s.} \rightarrow L_{g.s.} $ $ $ & \cite{28} & $ a=0 $ & $ a\neq0 $ & $ Exp.$ \cite{34}& $ Exp.$ \cite{36}\cr
			\mr 
			$ 11/2^{-}\rightarrow 7/2^{-} $  & 1.52 & 1.71 &1.70 & 2.03(33) & 1.75(12) \cr 
			$ 13/2^{-}\rightarrow 9/2^{-} $  & 1.83 & 2.19 &2.19 & 2.06(33) & 1.79(12) \cr 
			$ 15/2^{-}\rightarrow 11/2^{-} $ & 2.00 & 2.53 &2.52 & 2.31(36) & 2.00(13) \cr 
			$ 17/2^{-}\rightarrow 13/2^{-} $ & 2.11 & 2.80 &2.78 & 2.93(50) & 2.52(15) \cr 
			$ 19/2^{-}\rightarrow 15/2^{-} $ & 2.21 & 2.99 &2.96 & 3.21(50) & 2.82(19) \cr 
			$ 21/2^{-}\rightarrow 17/2^{-} $ & 2.25 & 3.18 &3.15 & 3.26(54) & 2.77(16) \cr 
			$ 23/2^{-}\rightarrow 19/2^{-} $ & 2.33 & 3.31 &3.27 & 3.37(70) & 2.89(6)  \cr
			\br
		\end{tabular}
	\end{indented}
\end{table}	

\begin{table}
	\caption{\label{tab8} The comparison of the theoretical predictions of $B(E_{2};L'_{g.s.} \rightarrow L_{g.s.})$ in units of $B(E_{2};9/2^-_{g.s.} \rightarrow 5/2^-_{g.s.})$ in two cases: $ a=0 $ and $ a\neq0 $ for $^{163}$Dy with those from \cite{28} and experimental values of \cite{34} and \cite{37}.}
	
	\begin{indented}
		\lineup
		\item[]\begin{tabular}{@{}*{6}{cccccc}}
			\br                              
			$ $ $ L'_{g.s.} \rightarrow L_{g.s.} $ $ $ &\cite{28}& $ a=0 $ & $ a\neq0 $ & $ Exp.$\cite{34}& $ Exp.$\cite{37}\cr
			\mr 
			$ 11/2^{-}\rightarrow 7/2^{-} $  & 1.54 & 1.70 & 1.67 & -        & 1.63(11) \cr 
			$ 13/2^{-}\rightarrow 9/2^{-} $  & 1.85 & 2.19 & 2.10 & 1.89(35) & 2.04(44) \cr 
			$ 15/2^{-}\rightarrow 11/2^{-} $ & 2.04 & 2.51 & 2.36 & 2.61(48) & 2.80(31) \cr 
			$ 17/2^{-}\rightarrow 13/2^{-} $ & 2.18 & 2.78 & 2.54 & 2.25(41) & 2.44(18) \cr 
			$ 19/2^{-}\rightarrow 15/2^{-} $ & 2.28 & 2.97 & 2.64 & 2.34(43) & 2.60(4)  \cr 
			$ 21/2^{-}\rightarrow 17/2^{-} $ & 2.38 & 3.15 & 2.71 & 2.25(41) & 2.44(24) \cr 
			$ 23/2^{-}\rightarrow 19/2^{-} $ & 2.45 & 3.28 & 2.75 & 2.07(38) & 2.28(35) \cr
			\br
		\end{tabular}
	\end{indented}
\end{table}

\begin{table}
	\caption{\label{tab9} The comparison of the theoretical predictions of $B(E_{2};L'_{g.s.} \rightarrow L_{g.s.})$ in units of $B(E_{2};9/2^+_{g.s.} \rightarrow 5/2^+_{g.s.})$ in two cases: $ a=0 $ and $ a\neq0 $ for $^{153}$Eu with those from \cite{29} and experimental values of \cite{38}.} 
	
	\begin{indented}
		\lineup
		\item[]\begin{tabular}{@{}*{6}{ccccc}}
			\br                              
			$ $ $ L'_{g.s.} \rightarrow L_{g.s.} $ $ $ & \cite{29} & $ a=0 $ & $ a\neq0 $ & $ Exp.$\cite{38}\cr
			\mr 
			$ 11/2^{+}\rightarrow 7/2^{+} $  & 1.55 & 1.70 & 1.70   & 1.50(7) \cr 
			$ 13/2^{+}\rightarrow 9/2^{+} $  & 1.86 & 2.22 & 2.21   & 1.63(4) \cr 
			$ 15/2^{+}\rightarrow 11/2^{+} $ & 2.07 & 2.56 & 2.54   & 1.92(4) \cr 
			$ 17/2^{+}\rightarrow 13/2^{+} $ & 2.22 & 2.87 & 2.85   & 1.82(6) \cr 
			$ 19/2^{+}\rightarrow 15/2^{+} $ & 1.88 & 3.11 & 3.06   & 1.98(9) \cr 
			$ 21/2^{+}\rightarrow 17/2^{+} $ & 2.05 & 3.32 & 3.28   & 2.02(6) \cr 
			$ 23/2^{+}\rightarrow 19/2^{+} $ & 2.07 & 3.55 & 3.46   & 1.97(8) \cr
			\br
		\end{tabular}
	\end{indented}
\end{table}

\section{Conclusion}\label{disc6}

In this work, we have studied the effect of DDMF when Coriolis contribution is included in the Bohr Hamiltonian for deformed odd mass nuclei. Our model is based on analytical results we have elaborated recently without taking into account the Coriolis interaction and using Davidson potential for $ \beta $ shape and the harmonic oscillator potential for the $ \gamma $ one. The mass coeffcients for the three collective mode motions are taken as different. Excited-state energies and E2 transition probabilities have been calculated for $^{173}$Yb, $^{163}$Dy, $^{155}$Eu and $^{153}$Eu nuclei and analysed within and outside the DDMF. So, compared to available experimental data, it is shown that the deformation-dependent mass coeffcients have a significant influence on the spectra of these nuclei with Coriolis contribution. We have found a good agreement with experimental data in this case. We have also invistigated the double effect of DDMF and the Coriolis force on the moments of inertia. We have found that their use conjointly has two impacts on the moments of inertia, namely: the Coriolis force causes staggering effect and the DDMF slows down the moments of inertia rates.


\section*{References}

\begin{thebibliography}{40}

\label{1} \bibitem[1]{1}  Bohr A 1952 {\it Mat. Fys. Medd. K. Dan. Vidensk. Selsk.} {\bf 26} 1
\label{2} \bibitem[2]{2}  Bohr A and Mottelson B R 1975 {\it Nuclear Structure Vol II: Nuclear Deformations} (New York: Benjamin)
\label{3} \bibitem[3]{3} Eisenberg J M and Greiner W 1970 {\it Nuclear Theory: Nuclear Models vol 1} (Amsterdam: North-Holland)
\label{4}\bibitem[4]{4} Davydov A S and Filipov A F 1958 {\it Nucl. Phys.} {\bf 8} 237
\label{5}\bibitem[5]{5} Davydov A S and Chaban A A 1960 {\it Nucl. Phys.} {\bf 20} 499
\label{6}\bibitem[6]{6} Davydov A S and Sardaryan R A 1962 {\it Nucl. Phys.} {\bf 37} 106
\label{7}\bibitem[7]{7} Pashkevich V V Sardaryan R A 1965 {\it Nucl. Phys.} {\bf 65} 401
\label{8}\bibitem[8]{8} Jean M 1960 {\it Nucl. Phys.} {\bf 21} 142
\label{9}\bibitem[9]{9} Fortunato L Vitturi A 2004 {\it J. Phys. G: Nucl. Part. Phys.} {\bf 30} 627
\label{10}\bibitem[10]{10} Fortunato L 2005 {\it Eur. Phys. J.} A {\bf 26} 1 
\label{11}\bibitem[11]{11} Caprio M A 2011 {\it Phys. Rev.} C {\bf 83} 064309
\label{12}\bibitem[12]{12} Ait Ben Hammou A, Chabab M, El Batoul A, Hamzavi M, Lahbas A, Moumene I and Oulne M 2019 {\it Eur. Phys. J. Plus} {\bf 134} 577
\label{13}\bibitem[13]{13} Bonatsos D, McCutchan E A, Minkov N, Casten R F, Yotov P, Lenis D, Petrellis D and Yigitoglu I 2007 {\it Phys. Rev.} C {\bf 76} 064312
\label{14}\bibitem[14]{14} Yigitoglu I and Bonatsos D 2011 {\it Phys. Rev.} C {\bf 83} 014303
\label{15}\bibitem[15]{15} Bonatsos D, Georgoudis P E, Lenis D, Minkov N and Quesne C 2011 {\it Phys. Rev.} C {\bf 83} 044321
\label{16}\bibitem[16]{16} Bonatsos D, Georgoudis P E, Minkov N, Petrellis D and Quesne C 2013 {\it Phys. Rev.} C {\bf 88} 034316
\label{17}\bibitem[17]{17} Chabab M, Lahbas A and Oulne M 2015 {\it Phys. Rev.} C {\bf 91} 064307
\label{18}\bibitem[18]{18} Bonatsos D, Minkov N and Petrellis D 2015 {\it J. Phys. G: Nucl. Part. Phys.} {\bf 42} 095104
\label{19}\bibitem[19]{19} Chabab M, El Batoul A, Lahbas A and Oulne M 2016 {\it J. Phys. G: Nucl. Part. Phys.} {\bf 43} 125107
\label{20}\bibitem[20]{20} Jolos R V and von Brentano P 2006 {\it Phys. Rev.} C {\bf 74} 064307
\label{21}\bibitem[21]{21} Jolos R V and von Brentano P 2007 {\it Phys. Rev.} C {\bf 76} 024309
\label{22}\bibitem[22]{22} Jolos R V and von Brentano P 2008 {\it Phys. Rev.} C {\bf 78} 064309
\label{23}\bibitem[23]{23} Ermamatov M J and Fraser P R 2011 {\it Phys. Rev.} C {\bf 84} 044321
\label{24}\bibitem[24]{24} Buganu P, Chabab M, El Batoul A, Lahbas A and Oulne M 2018 {\it Nucl. Phys.} A {\bf 970} 272
\label{25}\bibitem[25]{25} Chabab M, El Batoul A, Lahbas A and Oulne M 2016 {\it Physics Letters} B {\bf 758} 212
\label{26}\bibitem[26]{26} Chabab M, El Batoul A, Lahbas A and Oulne M 2018 {\it Annals of Physics} {\bf 392} 142
\label{27}\bibitem[27]{27} Chabab M, El Batoul A, El-ilali I, Lahbas A and Oulne M 2020 {\it Eur. Phys. J. Plus} {\bf 135} 201
\label{28}\bibitem[28]{28} Ermamatov M J, Srivastava P C, Fraser P R, Stránsky P and Morales I O 2012 {\it Phys. Rev.} C {\bf 85} 034307
\label{29}\bibitem[29]{29} Ermamatov M J, Yépez-Martínez H and Srivastava P C 2016 {\it  Pramana Journal of Physics} {\bf 86} 1055 
\label{30}\bibitem[30]{30} Ciftci H, Hall R L and Saad N 2003 {\it Journal of Physics} A {\it : Mathematical and General} {\bf 36} 11807
\label{31}\bibitem[31]{31} Davydov A S 1967 {\it Excited States of Atomic Nuclei}  (Moscow:Atomizdat)
\label{32}\bibitem[32]{32} Szego G 1939 {\it Orthagonal Polynomials} (New York: American Mathematical Society)
\label{33}\bibitem[33]{33} Cakirli R B, Blaum K and Casten R F 2010 {\it Phys. Rev.} C {\bf 82} 061304
\label{34}\bibitem[34]{34} Nuclear Data Sheets (http://nndc.bnl.gov/)
\label{35}\bibitem[35]{35} Wu X and Lei Y A 2008 {\it Chinese Physics} C {\bf 32} 112 
\label{36}\bibitem[36]{36} Oshima M, Matsuzaki M, Ichikawa S, Iimura H, Kusakari H,
Inamura T,Hashizume A and Sugawara M 1989 {\it Phys. Rev.} C {\bf 40} 2084
\label{37}\bibitem[37]{37} Oshima M, Minehara E, Kikuchi S, Inamura T, Hashizume A,
Kusakari H and Matsuzaki M 1989 {\it Phys. Rev.} C {\bf 39} 645 
\label{38}\bibitem[38]{38} Smith G F et al. 1998 {\it Phys. Rev.} C {\bf 58} 3171	
	
\end{thebibliography}
\end{document}